\documentclass[preprint,12pt]{elsarticle}

\usepackage{graphicx}
\usepackage{amssymb,amsmath,amsfonts}
\usepackage{bm}
\usepackage{natbib}
\usepackage{url}

\newdefinition{remark}{Remark}

\begin{document}

\begin{frontmatter}

\title{Hamiltonian Lorenz-like models}

\author[1]{Francesco Fedele}
\ead{fedele@gatech.edu}
\author[2]{Cristel Chandre}
\ead{cristel.chandre@cnrs.fr}
\author[3]{Martin Horvat}
\ead{martin.horvat@fmf.uni-lj.si}
\author[4]{Nedjeljka \v{Z}agar}
\ead{nedjeljka.zagar@uni-hamburg.de}

\affiliation[1]{organisation={School of Civil and Environmental Engineering, Georgia Institute of Technology, Atlanta, GA, 30332, USA}}
\affiliation[2]{organisation={CNRS, Aix Marseille Univ, I2M, 13009 Marseille, France}}
\affiliation[3]{organisation={Faculty of mathematics and physics, University of Ljubljana, Jadranska cesta 19, Ljubljana, Slovenia}}
\affiliation[4]{organisation={Center for Earth System Research and Sustainability, Meteorological Institute, Universität Hamburg, Bundesstrasse 53, 20146 Hamburg, Germany}}

\begin{abstract}
The reduced-complexity models developed by Edward Lorenz are widely used in atmospheric and climate sciences to study nonlinear aspect of dynamics and to demonstrate new methods for numerical weather prediction. A set of inviscid Lorenz models describing the dynamics of a single variable in a zonally-periodic domain, without dissipation and forcing, conserve energy but are not Hamiltonian. In this paper, we start from a general continuous parent fluid model, from which we derive a family of Hamiltonian Lorenz-like models through a symplectic discretization of the associated Poisson bracket that preserves the Jacobi identity. A symplectic-split integrator is also formulated. These Hamiltonian models conserve energy and maintain the nearest-neighbor couplings inherent in the original Lorenz model. As a corollary, we find that the Lorenz-96 model can be seen as a result of a poor discretization of a Poisson bracket. Hamiltonian Lorenz-like models offer promising alternatives to the original Lorenz models, especially for the qualitative representation of non-Gaussian weather extremes and wave interactions, which are key factors in understanding many phenomena of the climate system. 
\end{abstract}


\end{frontmatter}

\section{Introduction}

The Lorenz-96 is a forced dissipative system with quadratic nonlinearities.  While conceptually simple and computationally efficient, its appeal lies in its ability to replicate chaotic behavior through the quadratic term, which mimics advection, a fundamental process in the atmosphere and oceans. This chaotic nature underscores the challenges of weather forecasting, where small uncertainties in initial conditions can lead to significant forecast errors over time.  A common perception of the Lorenz-96 model dynamics is that of synoptic-scale, mid-latitude wave perturbations, even though the model does not account for rotation. The model has been proven valuable for studies of atmospheric predictability and uncertainties in weather forecasts~\citep[e.g.][]{Ott2004,TrevisanPancotti1998}, spatio-temporal chaos~\citep[e.g.][]{PazoSpatiotempchaos}, climate change \citep[e.g.][]{LucariniSarno2011} and as a tool for demonstrating new methods in data assimilation~\citep[e.g.][]{evensen2003ensemble,sakov2008impacts,fertigetal2007}.

The Lorenz-96 model yields rough and uncorrelated spatial fields~\citep{PazoSpatiotempchaos}. The re-designed model \citep{Lorenz2005} includes a correlation length and a suite of models with increasing complexity, which we refer to as the Lorenz-2005 models, characterized by smoother fields. In his discussion, Lorenz also mentioned that the proposed models are not derived from physical principles of dynamical balance. His purpose was to design the simplest chaotic models which share certain qualitative properties with the atmosphere, or at least with much larger physically-based atmospheric models~\citep{Lorenz2005}.

Both the inviscid Lorenz-96 and Lorenz-2005 models conserve energy, but none of them are Hamiltonian~\citep{LucariniBlenderLorenz96}. As a consequence, the Lorenz models produce hidden nonphysical dissipation~\citep[e.g.][]{Caligan2016}, which adds to and misleads the effects of the physical dissipation referred to as mechanical damping by Lorenz~\citep{Lorenz2005}. In general, it is desirable to have models whose inviscid, or ideal, part is Hamiltonian to avoid such nonphysical effects. For example, the Hamiltonian formulation of atmospheric dynamics provides an elegant framework for studying Rossby wave interactions \citep{Holm1986,McIntyreShepherd1987,VannesteVial1995}.

True to the essence of the original work by Lorenz~\citep{Lorenz2005}, we aim to design the simplest inviscid Lorenz-like models that are both chaotic and Hamiltonian. The paper is structured as follows: in Section~\ref{sec:Lorenz}, we introduce and briefly review the two Lorenz models. 

In Section~\ref{sec:methods}, we consider a general continuous parent model from which a family of Hamiltonian Lorenz-like  models is derived. In Section~\ref{sec:symplectic}, we formulate a symplectic-split integrator to numerically solve for the Hamiltonian Lorenz-like models. In Section~\ref{sec:example}, we investigate the properties of one model of the family with cubic Casimir invariant, and present a numerical investigation of its dynamics. The conclusions and outlook are presented in Section~\ref{sec:conclusion}. 

\section{The Lorenz-96 model}
\label{sec:Lorenz}

We consider the values  $\mathbf{X}=(X_0,X_1,\cdots,X_{N-1})$ of some scalar atmospheric quantity (e.g., temperature, pressure or vorticity) sampled at $N$ equally spaced points along a circle of constant latitude of the Earth, or grid points of a one-dimensional grid with periodic boundary conditions $X_{n+N} = X_n$, where $n=0,\ldots,N-1$ is a spatial index. The Lorenz-96 model is given by~\citep{Lorenz1996}:
%----------------------------------------
\begin{equation}
    \label{Lorenz96}
    \dot{X_n}= 
        X_{n-1} (-X_{n-2} + X_{n+1})  - X_n + F\,,
\end{equation}
%----------------------------------------
and includes nearest-neighbor couplings of the field variables. The linear term models dissipation, or mechanical damping~\citep{Lorenz2005}, and $F$ is a constant forcing. The Lorenz-96 model produces rough and uncorrelated spatial fields~\citep{PazoSpatiotempchaos}. To achieve smoother and more correlated fields, Lorenz designed a model with $K$ neighbor couplings in $2005$~\citep{Lorenz2005}, hereafter referred to as Lorenz-2005. This model is given by
%----------------------------------------
$$
    \dot{X_n} = 
        \left[{\mathbf{X}} , {\mathbf{X}}\right]_{n;K} - X_n + F \,,
$$
%----------------------------------------
where the Lorenz product is defined as
%----------------------------------------
$$
    [{\mathbf{X}} , {\mathbf{Y}}]_{n;K} = \frac{1}{K^2}
        \sum_{i,j=-J}^{J} \mathrm{w}_i \mathrm{w}_j
        \left(-X_{n-2K-i} Y_{n-K-j} + X_{n-K+j-i}Y_{n+K+j}\right),
$$
%----------------------------------------
with $J=\lfloor K/2 \rfloor$, where  $\lfloor x \rfloor$ is the greatest integer less than or equal to $x$. The weight coefficients $\mathrm{w}_j=1$ for $K$ odd and $\mathrm{w}_j = \left(\frac{1}{2}\,\mathrm{if}\,|j| = J,1\, \mathrm{otherwise}\right)$ for $K$ even.  Such an energy-preserving nonlinear term is a surrogate for nonlinear advection. The dynamics depends on the length $K$ of neighbor couplings and forcing strength $F$. For the smallest correlation length $K=1$, the Lorenz-2005 model reduces to the Lorenz-96 in Eq.~\eqref{Lorenz96}.  Both models are invariant under translation and chaotic~\citep[e.g.][]{MAIOCCHI,CarluGinelli}; that is, if $\mathbf{X}=(X_{n})_{n=0,\dots N-1}$ is a solution, then the cyclically shifted sequence $(X_{n+m})_{n=0,\dots N-1}$ is also a solution for any integer~$m$.  

In absence of forcing and dissipation (mechanical damping), the inviscid part of the Lorenz-96 model in Eq.~\eqref{Lorenz96} 
%----------------------------------------
\begin{equation}
    \label{inviscid96}
    \dot{X_n} = X_{n-1} (-X_{n-2} + X_{n+1})\,,
\end{equation}
%----------------------------------------
is conservative since the \emph{energy} 
%----------------------------------------
$$
    H=\frac{1}{2}\sum_{n=0}^{N-1} X_n^2\,,
$$
%----------------------------------------
is an invariant of motion. This suggests writing the inviscid Lorenz-96 in Eq.~\eqref{inviscid96} as
%----------------------------------------
$$
    \dot{X_n}=\{X_n, H\}\,,
$$
%----------------------------------------
where the antisymmetric bracket is defined as
%----------------------------------------
\begin{equation}
    \label{Lorenz96bracket}
    \{F,G\}=\nabla F \cdot {\mathbb J} \nabla G\,,
\end{equation}
%%----------------------------------------
with the gradient $\nabla F$ (with components $\partial F/\partial {X_n}$), ${\mathbb J}$ an $N\times N$ antisymmetric matrix whose coefficients are given by
%----------------------------------------
\begin{equation}
    \label{bracketcoeff}
    {\mathbb J}_{nm}=X_{n-1}\delta_{m, n+1} -X_{m-1}\delta_{n, m+1}\,,
\end{equation}
%%----------------------------------------
and $\delta_{n,m}$ is the Kronecker tensor. The bracket $\{\cdot , \cdot\}$ is bilinear, antisymmetric, i.e., $\{F , G \}=-\{G , F \}$, and satisfies the Leibniz rule, i.e., $\{F H , G \}= \{F , G \}H + F \{H , G \}$.
However, it was noted in \citep{LucariniBlenderLorenz96} that the Jacobi identity
%----------------------------------------
\begin{equation}\label{Jacobi}
    \{F , \{G,H\}\} + \{H , \{F,G\}\} +\{G , \{H,F\}\}=0\,,
\end{equation}
%----------------------------------------
is not satisfied. To show this, we consider the counterexample set by $F=X_{n+1}$, $G=X_n$, and $H=X_{n-2}$. In this case, we find that $\{F , \{G,H\}\} + \{H , \{F,G\}\} +\{G , \{H,F\}\}=-X_{n-3}\not= 0$. As a consequence, the inviscid Lorenz-96 model is not Hamiltonian. 

For the Lorenz-2005, the bracket is still that of the Lorenz-96 in Eq.~\eqref{Lorenz96bracket}, but the Poisson matrix is given by
$$
    {\mathbb J}_{nm}=\frac{1}{K^2}\sum_{i,j=-J}^J w_i w_j \left(X_{n-K-i+j}\delta_{m,n+K+j}-X_{m-K-i+j}\delta_{n,m+K+j} \right)\,.
$$
Here, setting $K=1$ (and hence $J=0$ and $w_j=1$) yields the Poisson matrix~\eqref{bracketcoeff} of Lorenz-96. Direct calculations show that the Jacobi identity is not satisfied for any $K$. So, none of the Lorenz models have a Hamiltonian inviscid part. All inviscid Lorenz models conserve energy, but produce hidden dissipation. This adds to the physical dissipation, which can result in misleading effects due to competing sources of dissipation~\citep{Caligan2016}. Ideally, models should have an inviscid counterpart that is Hamiltonian to prevent unphysical dissipative effects. Furthermore, if we believe that the Lorenz models can be derived from the first principles of fluid mechanics, it is expected that the reduced models preserve some basic structural properties of the parent model, specifically that their ideal part is Hamiltonian~\citep{Morrison1982,Salmon1988,Morrison1998,Morrison2006}.

\section{Methods and Lorenz-like models}
\label{sec:methods}

In this section, we construct inviscid Lorenz-like models with nearest neighbor coupling that are Hamiltonian. To achieve this, we start with a continuous model known to be Hamiltonian and discretize it in a manner that preserves the Jacobi identity. Below, we provide details of our method and derive a family of Hamiltonian Lorenz-like models.

\subsection{Discretization of one-dimensional Poisson brackets with one field variable}
\label{subsec:PB}

We consider a one-dimensional field~$u(x,t)$ and a generic Poisson fluid bracket~\cite{Morrison1998}
%------------------------------------------
\begin{equation}
    \label{eqn:PBau}
    \{\mathsf{F},\mathsf{G}\}=\int {\rm d}x\ a(u) \left(\mathsf{F}_u\partial_x \mathsf{G}_u - \mathsf{G}_u\partial_x \mathsf{F}_u\right),
\end{equation}
%------------------------------------------
where $a(u)$ is a generic function of the field variable and $\mathsf{F}_u=\frac{\delta \mathsf{F}}{\delta u}$ denotes the variational derivative of an observable $\mathsf{F}[u]$ with respect to~$u$. This noncanonical bracket acts on the algebra of scalar functionals of $u$ and is a Poisson bracket for all functions $a(u)$. It possesses a Casimir invariant defined by the functional
%------------------------------------------
$$
    \mathsf{C}[u]=\int {\rm d}x\ \phi(u(x))\,,
$$
%------------------------------------------
which Poisson-commutes with any observable, that is $\{\mathsf{C}[u],\mathsf{F}\}=0$ for all $\mathsf{F}[u]$, and where $\phi$ is the antiderivative of $1/{\sqrt{2 \vert a(u) \vert}}$. 

The objective of this section is to discretize the Poisson bracket~\eqref{eqn:PBau} for scalar functions of a discrete number of values of $u$ along the line, denoted by $X_n=u(x_n)$ for $n=0,\ldots,N-1$. This is a delicate procedure since any generic discretization of the derivatives in the bracket~\eqref{eqn:PBau} will most likely break the Jacobi identity in Eq.~\eqref{Jacobi}.  A good discretization is formulated as follows. 

Drawing on Ref.~\cite{Viscondi2017}, we first~\emph{beatify} the noncanonical bracket~\eqref{eqn:PBau} by eliminating (locally) the explicit dependence of the bracket on the field via a change of variables. 
To do so, we first consider an intermediate function $f(u)=\sqrt{2 \vert a(u) \vert}$ such that $f(u)\phi'(u)=1$. Then, the bracket~\eqref{eqn:PBau} is rewritten (locally) as
%------------------------------------------
\begin{equation}
    \label{PBbeatified}
    \{\mathsf{F},\mathsf{G}\}=\int {\rm d}x\ f(u) \mathsf{F}_u\partial_x \left[ f(u) \mathsf{G}_u\right]\,.
\end{equation}
%------------------------------------------
We can now perform the change of variables $v=\phi(u)$, which is locally invertible, and, from the identity $\mathsf{F}[u]=\overline{\mathsf{F}}[v]$, the bracket~\eqref{eqn:PBau} becomes
%------------------------------------------
\begin{equation}
    \label{eqn:PBb}
    \{\overline{\mathsf{F}},\overline{\mathsf{G}}\}=\int {\rm d}x\ \overline{\mathsf{F}}_v\partial_x \overline{\mathsf{G}}_v,
\end{equation}
%------------------------------------------
and acts in the algebra of functionals of $v$. This bracket is~\emph{beatified}~\citep{Viscondi2017}, in the sense that it does not have any explicit dependence on the new field variable $v$ and, therefore, it is already in a form suitable for discretization. In other words, any discretization will yield a Poisson bracket, provided that the bracket is antisymmetric, lacks explicit dependence on the field variables, and involves only first-order derivatives.

Specifically, we construct a Poisson bracket for the functions $\overline{F}$ of the discrete variables $Y_n=v(x_n)$ for $n=0,\ldots,N-1$ as follows: we first discretize $\overline{\mathsf{G}}_v$ as $\overline{G}_{Y_n}$, and the first-order derivatives $\partial_x \overline{\mathsf{G}}_v$ as a sum of centered finite differences, i.e.,~$\sum_k \xi_k\left(\overline{G}_{Y_{n+k}}-\overline{G}_{Y_{n-k}}\right)$, where $\overline{G}$ is a function of the variables~$Y_n$ and $\overline{G}_{Y_n}=\partial \overline{G}/\partial {Y_n}$~(for $n=0,\ldots, N-1$). This yields the Poisson bracket
%------------------------------------------
\begin{equation}
    \label{eqn:PBdy}
    \{\overline{F},\overline{G}\}=\sum_{n=0}^{N-1} \overline{F}_{Y_n}\sum_{k=1}^{K} \xi_k \left(\overline{G}_{Y_{n+k}}-\overline{G}_{Y_{n-k}} \right) \,,
\end{equation}
%------------------------------------------
where the constants $\xi_k$ are arbitrary. At each node~$n$, the neighbor coupling extends to $K$ nodes on each side. The bracket~\eqref{eqn:PBdy} is Poisson for any values of the coefficients $\xi_k$, which are thus  arbitrary. If the interest is to guarantee also a symmetric finite-difference discretization of $\partial_x$ to order~$K$, which involves sites at $n+k$ and $n-k$, but not at $n$, then we must impose $K$ further constraints, i.e., $\sum_{k=1}^K 2 k\xi_k = 1$, $\sum_{k=1}^K k^2\xi_k = 0$, etc...~\citep[e.g.][]{horvat2012computational}. In what follows, we will not restrict these coefficients to allow for more generality.

The primary reason the bracket~\eqref{eqn:PBdy} satisfies the Jacobi identity is that it does not explicitly depend on the variables $Y_n$. Next we transform back to the coordinates $X_n$, and the Poisson bracket acting in the algebra of scalar functions $F$ of $X_n=u(x_n)$ becomes 
%------------------------------------------
\begin{equation}
    \label{newbracket}
    \{F,G\}=\sum_{n=0}^{N-1} f(X_n)F_{X_n}\sum_{k=1}^{K} \xi_k \left[ f(X_{n+k})G_{X_{n+k}}- f(X_{n-k})G_{X_{n-k}} \right] \,,
\end{equation}
%------------------------------------------
where $F_{X_n}=\partial F/\partial {X_n}$. As expected, this noncanonical Poisson bracket has the Casimir invariant
%------------------------------------------
$$
    C(\mathbf{X})= \sum_n \phi(X_n)\,.
$$
%------------------------------------------
The reduction procedure from bracket~\eqref{eqn:PBau} to bracket~\eqref{newbracket} not only preserves the crucial Poisson property, but also retains a Casimir invariant similar to that of the parent model.

If $N$ is not a multiple of $K+1$, $C$ is the only Casimir invariant. Otherwise, if $N$ is a multiple of $K+1$, there might be more Casimir invariants, depending on the value of the coefficients $\xi_k$. To see this, we divide the set of indices $n=0,\ldots,N-1$ into $K+1$ sets~${\cal G}_k$, for $k=0,\ldots,K$: An index~$n$ belongs to the set~${\cal G}_k$ if and only if~$n=k \mbox{ mod } (K+1)$. 
For instance, if $K=1$, there are two sets:  the set~${\cal G}_0$ of even integers and ${\cal G}_1$ of odd integers. 

We use two rather obvious statements: $(i)$ ${\cal G}_{m+k \mbox{ mod } (K+1) } \cap {\cal G}_k= \varnothing$ for all $m=1,\ldots,K$. $(ii)$ For all $l\neq k$, there exists a unique index $m^*(l,k)$ such that ${\cal G}_{m^* + l\mbox{ mod } (K+1)} = {\cal G}_k$. 
Next, we consider the observables
$$
    C_k({\bf X}) = \sum_{n\in {\cal G}_k} \phi(X_n)\,,
$$
and $C=\sum_k C_k$. We can now compute~$\{X_n,C_k\}$ for all $n$'s:
$$
    \{X_n, C_k\}=f(X_n)(\xi_{m^*}-\xi_{K+1-m^*})\,.
$$
If the coefficients~$\xi_k$ satisfy $\xi_m=\xi_{K+1-m}$ for all $m=1,\ldots,K$, then $C_k$ is a Casimir invariant for all $k=0,\ldots,K$. This identity is automatically satisfied for $K=1$. This is because, for $m=1,\cdots,N$, we have $m^*=1$, and thus~$K+1-m^*=m^*$. 

In summary, if $N$ is a multiple of $K+1$, and if $\xi_k=\xi_{K+1-k}$ for $k=1,\ldots,K$, there are the $K+1$ Casimir invariants $C_k$ defined above.  In all the other cases, there is a single Casimir invariant $C$.

\begin{remark}[Foliation and hidden dissipation] The existence of Casimir invariants foliates the state space into symplectic leaves on which the dynamics occurs. If a reduced bracket, such as the one for the inviscid Lorenz models~\citep{Lorenz2005}, does not satisfy the Jacobi identity, the Casimir invariants are not conserved. As a consequence, the flow \emph{dissipates} across the symplectic leaves, drastically affecting the resulting dynamical properties~\citep{Caligan2016}.
\end{remark}

\subsection{Hamiltonian Lorenz-like models}

Equation~\eqref{newbracket} defines a family of Poisson brackets in the variables $X_n$. To fully specify a dynamical model, a Hamiltonian is required. Similar to the original Lorenz models, we consider the quadratic energy
\begin{equation}
    \label{eqn:Ham}
    H(\mathbf{X})=\frac{1}{2}\sum_{n=0}^{N-1} X_n^2\,,
\end{equation}
and ensure periodic boundary conditions, i.e., $X_{n+N}=X_n$ for all $n$.
The equations of motion follow from the Poisson bracket~\eqref{newbracket}, i.e., $\dot{X}_n=\{X_n,H\}$:
\begin{equation}
    \label{eqn:eqnmot}
    \dot{X}_n=f(X_n) \sum_{k=1}^K\xi_k \left( X_{n+k}f(X_{n+k})- X_{n-k}f(X_{n-k})\right)\,.
\end{equation}
This is a family of Hamiltonian Lorenz-like models defined by the generic function $f(X)$, or alternatively by the function $\phi(X)$ such that~$f(X)\phi'(X)=1$. For $K>1$, they resemble the Lorenz-2005 models with $K$-range coupling~\citep{Lorenz2005}. 

For $K=1$, we obtain a family of Hamiltonian models similar to the Lorenz-96 with nearest neighbor coupling as
\begin{equation}
    \label{eqn:eqnmot96}
    \dot{X}_n=f(X_n)\left( X_{n+1}f(X_{n+1})- X_{n-1}f(X_{n-1})\right)\,,
\end{equation}
where we have set $\xi_1=1$ and the Casimir invariant
$$
    C(\mathbf{X})=\sum_n \phi(X_n)\,.
$$

The transformation~$Y_n=\phi(X_n)$ makes the Casimir invariant linear in the variables $Y_n$, but the transformed Hamiltonian is no longer quadratic
$$
    \overline{H}({\bf Y})=\frac{1}{2}\sum_n \left(\phi^{-1}(Y_n)\right)^2\,,
$$
unless $\phi$ is linear. The associated Poisson bracket follows from Eq.~\eqref{eqn:PBdy} with $K=1$ and $\xi_k=1$ as 
\begin{equation}
    \label{eqn:PBdy2}
    \{\overline{F},\overline{G}\}=\sum_{n=0}^{N-1} \overline{F}_{Y_n} \left(\overline{G}_{Y_{n+1}}-\overline{G}_{Y_{n-1}} \right) \,,
\end{equation}
and the dynamical equations
$$
    \dot{Y}_n=\{Y_n,\overline{H}\}=\overline{H}_{Y_{n+1}}-\overline{H}_{Y_{n-1}}\,,
$$
or explicitly 
%------------------------------------------
\begin{equation}
    \label{Ynmodel}
    \dot{Y}_n=\phi^{-1}(Y_{n+1})\partial_{Y_{n+1}} \phi^{-1}(Y_{n+1})- \phi^{-1}(Y_{n-1})\partial_{Y_{n-1}} \phi^{-1}(Y_{n-1})\,,
\end{equation}
%------------------------------------------
and the Casimir invariant~$\overline{C}(\mathbf{Y})=\sum_n Y_n$ resembles the linear momentum. 

\paragraph*{Continuous parent models}
The continuous parent model of the discrete field~$\mathbf{X}$ is derived from the Poisson bracket~\eqref{PBbeatified} using the quadratic Hamiltonian~$\mathsf{H}[u]=\int {\rm d}x\ u(x)^2/2$ as
%------------------------------------------
\begin{equation}
    \label{parentmodel}
    \partial_t u=\{u,\mathsf{H}\}=f(u)\partial_x \left[ f(u) u\right]\,, 
\end{equation}
%------------------------------------------
and the Casimir invariant~$\mathsf{\mathsf{C}}[u]=\int {\rm d}x\ \phi(u(x))$. In contrast, the transformed field~$v(x)=\phi(u(x))$, continuous counterpart of the discrete field~$\mathbf{Y}$, evolves according to the Poisson bracket~\eqref{eqn:PBb} with the non-quadratic Hamiltonian
%------------------------------------------
$$
\overline{\mathsf{H}}[v]=\frac{1}{2}\int {\rm d}x\ \left( \phi^{-1}(v(x))\right)^2\,.
$$
%------------------------------------------
The associated dynamical equations are given by 
$$
    \partial_t v=\partial_x \left(\phi^{-1}(v) \partial_v{\phi^{-1}(v)}\right)\,, 
$$
%------------------------------------------
with the linear Casimir invariant~$\overline{\mathsf{C}}[v]=\int {\rm d}x\ v(x)$.  

\begin{remark}
The two discrete sets of variables, $X_n$ and $Y_n$, or their continuous counterparts $u$ and $v$, are related by the transformation~$Y_n=\phi(X_n)$, or $v=\phi(u)$. The Hamiltonian is quadratic in the $u$ or $X_n$~variables, while the Casimir invariant is nonlinear. Conversely, the Casimir invariant is linear in the $v$ or $Y_n$ variable, but in this case, the Hamiltonian is not quadratic.
\end{remark}

\begin{remark}[\emph{Bad} bracket discretization] Most discretizations of the Poisson bracket~\eqref{eqn:PBau} will yield brackets in the discrete variables $X_n$ that fail the Jacobi identity and introduce fake dissipation in the dynamics. For example, the Lorenz-96 model can be seen as following from a bad discretization of the Poisson bracket~\eqref{eqn:PBau} with $a(u)=u$. For instance, we discretize~$a(u)=u$ as $X_{n-1}$, $\mathsf{F}_u$ as $F_{X_n}$, and the first-order derivatives $\partial_x \mathsf{G}_u$ as~$G_{X_{n+1}}-G_{X_n}$, where $F_{X_n}=\partial F/\partial X_n$. As a result, we obtain the discrete bracket
\begin{equation}
    \label{br}
    \{F,G\}=\sum_{n=0}^{N-1} X_{n-1} \left(F_{X_n} G_{X_{n+1}}-G_{X_n}F_{X_{n+1}} \right)\, ,
\end{equation}
which is the same bracket as the one defined by the coefficients in Eq.~\eqref{bracketcoeff}.
Using the energy $H=\sum_n X_n^2/2$ yields the Lorenz-96 model
$$
    \dot{X_n}=\{X_n,H\}=X_{n-1}(X_{n+1}-X_{n-2})\,.
$$
This conserves the energy~$H$, but it is not Hamiltonian since the bracket~\eqref{br} is not Poisson, in the sense that the Jacobi identity is not satisfied~(see also~\citep{LucariniBlenderLorenz96}). As a result, a hidden dissipation may result, misleading the effects of any added physical (mechanical) damping~\citep{Caligan2016}.    
\end{remark}

\begin{remark}[Linear model]
The model of the Hamiltonian family~\eqref{eqn:eqnmot96} that has the linear momentum $M=\sum_n X_n$ as a Casimir invariant is defined by choosing~$\phi(X)=X/U$ and $f(X)=1/\phi'(X)=U$, that is
$$
    \dot{X_n}=U(X_{n+1}-X_{n-1})\,,
$$
which approximates linear advection at constant speed~$U$. 
\end{remark}

\begin{remark}[A seemingly innocent model]
The simplest nontrivial choice for $f$ is a linear function $f(X)=a+b X$, which yields the model 
$$
    \dot{X}_n=(a+b X_n)\left( X_{n+1}(a+ b X_{n+1})- X_{n-1}(a + b X_{n-1})\right)\,. 
$$ 
At first glance, the simplicity of this model is reminiscent of the original Lorenz-96. 
However, despite its simplicity, the Hamiltonian model generates non-trivial flow dynamics, as explained below. 
 
The associated Casimir invariant is
$$
    C=\sum_n \phi(X_n)=\sum_n \ln{|a+b X_n|}\,,
$$
where $\phi(X)=\int {\rm d}x\ f(X)^{-1}=\ln{|a+b X|}$, or equivalently $\exp(C)=\prod_n (a+b X_n)$. The signs $\mathrm{sgn}(a+b X_n)$ are also invariant under the dynamics. This is because the Casimir invariant is infinite on the hyperplanes $X_n=-a/b$, preventing the dynamical flow from crossing them. These barriers partition the phase space into disconnected regions, each with its own distinct dynamical and ergodic properties. 
Note that the inviscid Lorenz-96 model has no conserved quantity besides the energy, so the motion is not restricted to disconnected components in phase space. 
\end{remark}

\section{Symplectic-split integration}
\label{sec:symplectic}

The family of Hamiltonian Lorenz-like models in Eq.~\eqref{eqn:eqnmot} lends itself very naturally to a symplectic integration when $N$ is a multiple of $K+1$, where $K$ is the number of interacting sites on each side of a given $X_n$. Split the Hamiltonian~\eqref{eqn:Ham} into $K+1$ components as
$$
    H=\sum_{k=0}^K H_k\,,  
$$
where
$$
    H_k=\frac{1}{2}\sum_{n\in {\cal G}_k} X_n^2\,.
$$
Then, the Liouville operator ${\cal L}=\{\cdot,H\}$ is the sum of $K+1$ Liouville operators ${\cal L}=\sum_k{\cal L}_k$ where ${\cal L}_k=\{\cdot,H_k\}$. A particularly useful case corresponds to $K=1$ where the split is performed over odd and even indices. The time-propagators $\mathrm{e}^{t {\cal L}_k}$ can be explicitly computed. 
From the properties $(i)$ and $(ii)$ of the sets ${\cal G}_k$ in section~\ref{subsec:PB}, we deduce the equations of motion associated with $H_k$:
\begin{align*} 
    & \dot{X}_{n} \equiv {\cal L}_k X_n=0 \qquad \mbox{ for } n\in {\cal G}_k, \\
    & \dot{X}_{n} = f(X_{n}) \left[\xi_{m^*(l,k)} X_{n+m^*(l,k)} f(X_{n+m^*(l,k)})\right.\\
    & \left. \qquad \qquad - \xi_{K+1-m^*(l,k)} X_{n+m^*(l,k)-K-1} f(X_{n+m^*(l,k)-K-1})  \right] \qquad \mbox{ for } n\notin {\cal G}_k,
\end{align*}
for $l$ defined by $n=l \mbox{ mod} (K+1)$. In the second equation, we see that only variables with indices in ${\cal G}_k$ are involved in the right-hand side, and we notice that these variables are constant according to the first equation above.
In other words, the time propagator $\mathrm{e}^{t {\cal L}_k}$ leaves the variables $X_n$ with $n\in {\cal G}_k$ unchanged, and the other variables satisfy the following uncoupled ordinary differential equations:
$$
    \dot{X}_{n} =K_{n} f(X_{n}) \qquad  \mbox{ for } n\notin {\cal G}_k,
$$
where 
\begin{eqnarray}
    K_n&=&\xi_{m^*(l,k)} X_{n+m^*(l,k)} f(X_{n+m^*(l,k)})\nonumber\\
    && - \xi_{K+1-m^*(l,k)} X_{n+m^*(l,k)-K-1} f(X_{n+m^*(l,k)-K-1}). \label{eqn:Kn}
\end{eqnarray}
The solution is given by
%------------------------------
$$
    X_{n}(t)=\phi^{-1}\left( K_{n} t + \phi(X_{n}(0)  \right)\,,
$$
%------------------------------
where $\phi^{-1}$ is the inverse function of $\phi$, for $n\notin {\cal G}_k$.
Consequently, the time propagation from $t$ to $t+h$ associated with $H_k$ is explicitly computed as
\begin{align*}
    & X_{n}(t+h)=X_{n}(t) \qquad \mbox{ for } n\in {\cal G}_k,\\
    & X_{n}(t+h)=\phi^{-1}\left(K_n h + \phi(X_{n}(t)) \right) \qquad \mbox{ for } n\notin {\cal G}_k\,.
\end{align*}
Since the split is based on the same Poisson bracket~\eqref{newbracket}, all Casimir invariants of this bracket are preserved by each time propagator~$\mathrm{e}^{t {\cal L}_k}$. As a result, the split integration remains symplectic on each symplectic leaf defined by the values of the Casimir invariants.
Given that the split is constructed with the same Poisson bracket~\eqref{newbracket}, all its Casimir invariants are preserved by each of the time propagators $\mathrm{e}^{t {\cal L}_k}$. 

Drawing on Ref.~\citep{McLachlan2022}, we can approximate the dynamics as given by the product of simpler time-step propagators as
%------------------------------------------
\begin{equation}
    \label{SI}
   \mathrm{e}^{h {\cal L}}=\chi(\alpha_s h)\chi^*(\alpha_{s-1} h)\cdots \chi(\alpha_2 h)\chi^*(\alpha_1 h) +O(h^{p+1})\,, 
\end{equation}
%------------------------------------------
where 
%------------------------------------------
\begin{eqnarray*}
    \chi(h)&=&{\rm e}^{h{\cal L}_{K}}{\rm e}^{h{\cal L}_{K-1}}\cdots {\rm e}^{h{\cal L}_{0}},\\
    \chi^*(h)&=&{\rm e}^{h{\cal L}_{0}}{\rm e}^{h{\cal L}_{1}}\cdots {\rm e}^{h{\cal L}_{K}}\,.
\end{eqnarray*}
%------------------------------------------
Here, $s$ denotes the number of steps used in the scheme, and $p$ denotes the order of the symplectic scheme. The coefficients $\alpha_j$ are determined using the Baker–Campbell–Hausdorff formula and possibly optimized according to the scheme used in the computation~\citep{McLachlan2022}. For instance, the simplest symplectic scheme is the first order~($p=1$) symplectic Euler integrator. A second order~($p=2$), the Verlet symplectic integrator, has coefficients $\alpha_1=\alpha_2=1/2$. For the numerical simulations, we use a more accurate integration scheme, the optimized BM4 symplectic scheme~\citep{Blanes_2002} which is of fourth order~($p=4$). We refer to~\cite{pyHamSys2024} for a numerical implementation of these symplectic integrators in Python. 

\begin{remark}[Lorenz symplectic maps] The symplectic integration of Hamiltonian Lorenz-like models yields a method for constructing $N$-dimensional symplectic maps. For example, using the Euler integrator~($p=1$) in Eq.~\eqref{SI} yields the map
$$
     X_{n}^\prime = \phi^{-1}\left(K_{n}h +\phi(X_{n}) \right)\,,
$$
where the coupling between the variables with odd and even indices is hidden in $K_n$ given by Eq.~\eqref{eqn:Kn}. These symplectic maps have the same Casimir invariants of the Poisson bracket~\eqref{newbracket}, that is, $C=\sum_n \phi(X_n)$. 

Another symplectic map can be derived for the field $Y_n=\phi(X_n)$ leading to
$$
     Y_{n}^\prime = K_n h + Y_n\,,
$$
where
\begin{eqnarray*}
    K_n&=&\xi_{m^*(l,k)} \phi^{-1}(Y_{n+m^*(l,k)}) f(\phi^{-1}(Y_{n+m^*(l,k)}))\\
    && - \xi_{K+1-m^*(l,k)} \phi^{-1}(Y_{n+m^*(l,k)-K-1}) f(\phi^{-1}(Y_{n+m^*(l,k)-K-1}))\,,
\end{eqnarray*}
with $n=l \mbox{ mod } (K+1)$. The two symplectic maps for the $X_n$ and $Y_n$ fields naturally lend themselves to define feed-forward neural networks~\citep[e.g.][]{SympNet20201,BurbyHenonNet2021}.
\end{remark}

\section{Hamiltonian Lorenz-like models with cubic Casimir invariant}
\label{sec:example}

\subsection{The model and some of its properties}

We choose a Hamiltonian Lorenz-like model of the family in Eq.~\eqref{eqn:eqnmot2} with $K=1$ and a cubic Casimir invariant
$$
    C=\sum_n \left(X_n + \alpha X_n^2 +\beta X_n^3\right)\,,
$$
i.e., $\phi(X)=X+\alpha X^2 + \beta X^3$. %where we have re-scaled time to set the coefficient of~$X_n$ equal to $1$.
The associated function $f$ is given by 
\begin{equation}
    \label{FX2}
    f(X)=\frac{1}{\phi'(X)} = \frac{1}{1+2\alpha X + 3\beta X^2}\,, 
\end{equation}
which is always bounded for $\beta> \alpha^2/3$.  The Hamiltonian Lorenz-like model that follows from Eq.~\eqref{eqn:eqnmot} is 
%-----------------------------------------
\begin{equation}
    \label{eqn:eqnmot2}
    \dot{X}_n=\frac{1}{1+2\alpha X_n + 3\beta X_n^2} \left(  \frac{X_{n+1}}{1+2\alpha X_{n+1} + 3\beta X_{n+1}^2}- \frac{X_{n-1}}{1+2\alpha X_{n-1} + 3\beta X_{n-1}^2}\right).
\end{equation}
%-----------------------------------------
Besides the Hamiltonian $H=\sum_n X_n^2/2$, the dynamics conserves the Casimir invariant $C$. As mentioned above, for $N$ even, this Casimir invariant is the sum of two Casimir invariants: $C_0=\sum_{n~{\rm even}} \phi(X_n)$ and $C_1=\sum_{n~{\rm odd}} \phi(X_n)$. 
A linear model is recovered for $\alpha=0$ and $\beta=0$ with the linear momentum $M=\sum_n X_n$ as the Casimir invariant. 

\paragraph*{Symplectic integration}
In the symplectic-split integration scheme, the computation of $\phi^{-1}$ is required. For our choice of $\phi$, this is always well defined since the function $\phi$ is a third-order polynomial which is strictly monotonous, since~$\phi'(X)=1/f(X)>0$. 

\paragraph*{Uniform equilibria}
For each value of the energy $E$, there exists a unique uniform equilibrium ${\mathbf{X}}_c =\{X_n=c\}_{n=0}^{N-1}=\{c,c,\cdots,c,c\}$, with $c=\pm \sqrt{2 E/N}$. These equilibria are on the symplectic leaf defined by the Casimir invariant $C=N \phi(c)$. These equilibria are marginally stable for any energy $E$. 

\paragraph*{Zig-zag equilibria}
For $N$ even, there exist $\emph{zig-zag}$ equilibria 
${\mathbf{X}}_{a_0 a_1} =\{a_0,a_1,\cdots,a_0,a_1\}$, where the values of $a_0$, $a_1$ are set by the Casimir invariants $C_0$ and $C_1$, and follow by solving the cubic equations
%-----------------------------------------
\begin{equation}
    \label{phieq}
    \phi(a_j)- 2C_j/N=0\,,\quad j=0,1\,,  
\end{equation}
%-----------------------------------------
each of which admits only one real root since $\phi$ is a monotonous function with range ${\mathbb R}$. The energy of this equilibrium is $E=N(a_0^2+a_1^2)/4$. 

The equilibria are unstable~(hyperbolic) if $S=(1 - 3\beta  a_0^2)(1 - 3\beta a_1^2)<0$, otherwise they are marginal~$(S\ge 0)$. More generally, the condition is $f(a_0)f(a_1)\left(a_0f'(a_0)+f(a_0)\right)\left(a_1f'(a_1)+f(a_1)\right)<0$ for the equilibria to be unstable. Otherwise, these equilibria are marginal. 
 
The uniform equilibria are special cases with $a_0=a_1=c$ and are marginal since $S=f(c)^2\left(c f'(c)+f(c)\right)^2\ge 0$ for any real $c$.

\paragraph*{Quartic model for weak fields}
The dynamics of the transformed variables $Y_n=\phi(X_n)=X_n + \alpha X_n^2 + \beta X_n^3$ can be truncated to that of a quartic Hamiltonian model when the field is weak, i.e.,  $Y_n\ll 1$. First, we invert the relation $Y_n=\phi(X_n)$ as
$$
    X_n=\phi^{-1}(Y_n) = Y_n - \alpha Y_n^2 + (2 \alpha^2 - \beta)  Y_n^3\ +O(Y_n^4)\,,
$$
and the Hamiltonian is truncated as
$$
    H=\sum_n \frac{X_n^2}{2}=\sum_n \left(\frac{1}{2}Y_n^2 -\alpha Y_n^3 + \frac{5\alpha^2-2\beta}{2} Y_n^4\right) + O(Y_n^5)\,.
$$
The Casimir invariant $C=\sum_n Y_n$ is the linear momentum. The associated Poisson bracket is given in Eq.~\eqref{eqn:PBdy2}
and the dynamical equations
$$
    \dot{Y}_n=\{Y_n,H\}=H_{Y_{n+1}} -H_{Y_{n-1}}\,,
$$
or explicitly
%-----------------------------------------
\begin{equation}
    \label{Lorenzlikemodel7}
    \dot{Y}_n=Y_{n+1}-Y_{n-1} -3\alpha (Y_{n+1}^2-Y_{n-1}^2) + 2(5\alpha^2-2\beta) (Y_{n+1}^3-Y_{n+1}^3)\,.
\end{equation}
%-----------------------------------------
 
The continuous model has the linear Casimir invariant $\mathsf{C}[v]=\int {\rm d} x\,v(x) $ and the quartic Hamiltonian
$$
    \mathsf{H}=\frac{1}{2}\int {\rm d} x\, \left(\phi^{-1}(v)\right)^2=\int \left(\frac{1}{2}v^2 -\alpha v^3 + \frac{5\alpha^2-2\beta}{2}v^4\right)\,,
$$
where we have neglected the $O(v^5)$ terms. The associated Poisson bracket is given by Eq.~\eqref{eqn:PBb}, and the dynamical equation
%--------------------------------------------
$$
    \partial_t v=\{\mathsf{v},\mathsf{\mathsf{H}}\}= \partial_x v -6\alpha v\partial_x v + 6(5\alpha^2-2\beta) v^2\partial_x v\,,
$$
%--------------------------------------------
is a dispersionless modified KdV, which is integrable. The discrete model in Eq.~\eqref{Lorenzlikemodel7} is likely not integrable, though. 

%--------------------------------------
%--------------------------------------
\begin{figure}
    \centering
    \includegraphics[width=1.1\textwidth]{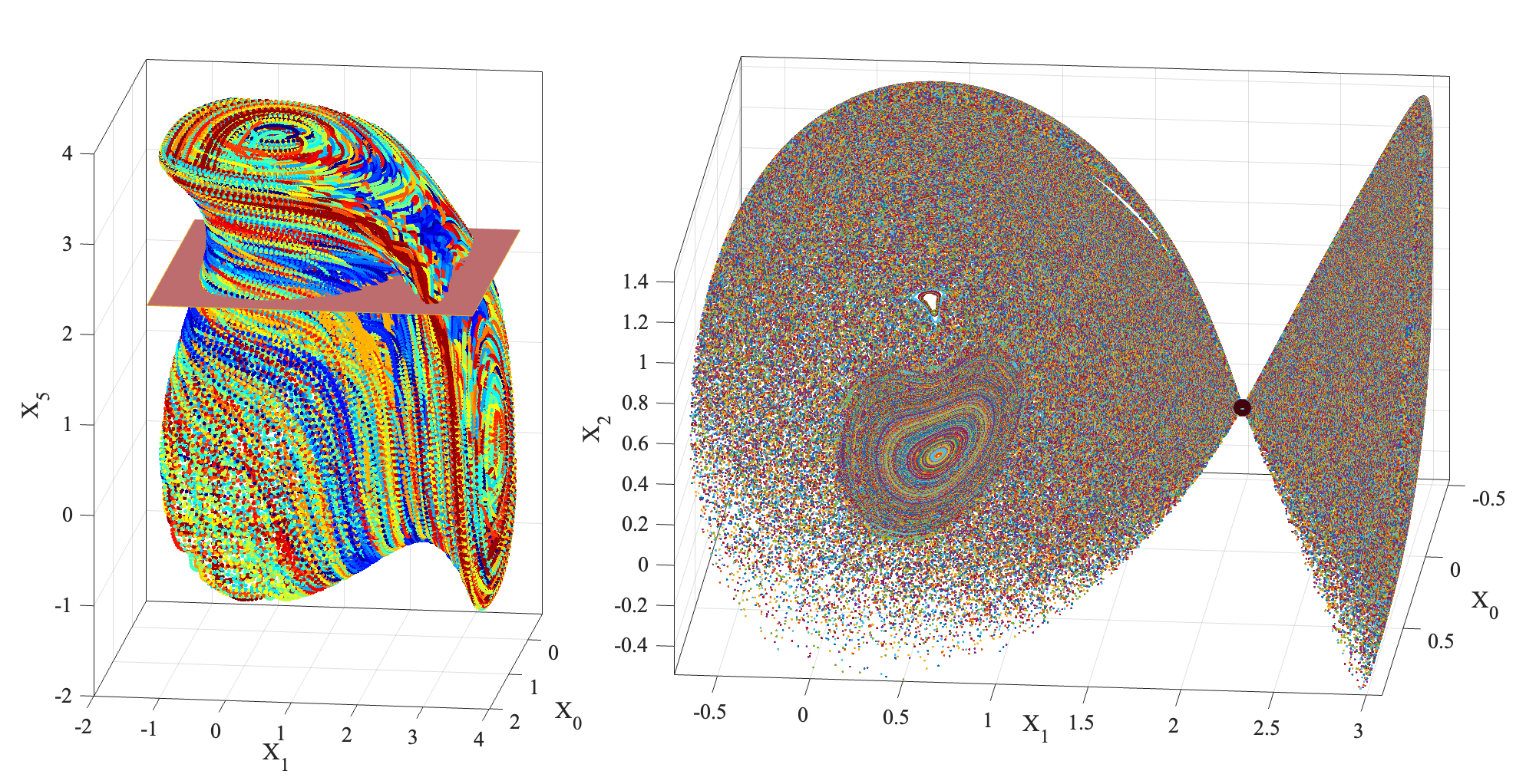}
    \caption{Dynamics in the region of the zig-zag unstable equilibrium ${\bf X}_{a_0 a_1}$ of the Hamiltonian Lorenz-like model given by Eq.~\eqref{eqn:eqnmot2} for $N=6$. The Casimir invariants are $C_0=2$ and $C_1=18$, and Hamiltonian $H=E\approx 8.074$. (Left panel) Some sample trajectories projected in $(X_0,X_1,X_5)$.  
    (Right panel) Poincar\'e section in $(X_0,X_1,X_2)$ of some sample trajectories. The Poincar\'e section is done at~$X_5=a_1$ with $\dot{X}_5<0$, and shown on the left panel as an orange plane. The unstable zig-zag fixed point is depicted with a black dot. } 
    \label{Fig1}
\end{figure}
%--------------------------------------
%--------------------------------------

\subsection{Numerical results}

Here, we consider the one-parameter family of Hamiltonian Lorenz models described in Eq.~\eqref{eqn:eqnmot2}, specifically with the function $f(X)=1/(1+X^2)$ and parameters $\alpha=0$ and $\beta=1/3$, as specified in Eq.~\eqref{FX2}. 

\subsubsection{Low~$N$}

First, we examine the chaotic nature of the flow with~$N=6$. This flow has two Casimir invariants~$C_0$ and $C_1$. Thus, the symplectic leaves are four-dimensional and the flow can be easily visualized, for instance, using Poincar\'e sections, since the model has only two effective degrees of freedom. 

The left panel of Fig.~\ref{Fig1} depicts the flow dynamics around the zig-zag unstable equilibrium point~${\mathbf{X}}_{a_0 a_1}=\{a_0,a_1,a_0,a_1,a_0,a_1\}$. Here,  $a_0\approx 0.596$ and $a_1\approx 2.242$ are set by the two Casimir invariants $C_0=18$ and $C_1=2$ using Eq.~\eqref{phieq}, and hence, the Hamiltonian value is $E\approx 8.074$. 
The Lyapunov spectrum, which consists of a pair of contracting and expanding exponents $(\lambda,1/\lambda)$, $\lambda\approx 0.062$, along with four marginal exponents. 
The Hamiltonian Lorenz-like flow compared to the Lorenz-96 flow also exhibit chaotic dynamics, with one noticeable difference being that the contracting directions are paired with the expanding ones~\citep{chaosbook}.

To better visualize the chaotic nature of the flow, we set a Poincaré section at~$X_5=a_1$ with~$\dot{X}_5<0$, passing through the unstable \emph{zig-zag}~equilibrium~${\mathbf{X}}_{a_0 a_1}$. The center panel of Fig.~\ref{Fig1} shows the associated Poincaré section in the $(X_0, X_1, X_2)$ space, which is a curved surface. In addition to the large elliptic island on the left, a tiny blue island and a yellow elongated strip-like elliptic island are also present. 
As expected in typical Hamiltonian systems, the phase space is mixed, composed of chaotic and regular~(elliptic) regions. The Poincar\'e sections reveal large chaotic regions surrounding smaller elliptic islands, which form around elliptic periodic orbits. 

\subsubsection{Large~$N$}

\begin{figure}
    \centering
    \includegraphics[width=1\textwidth]{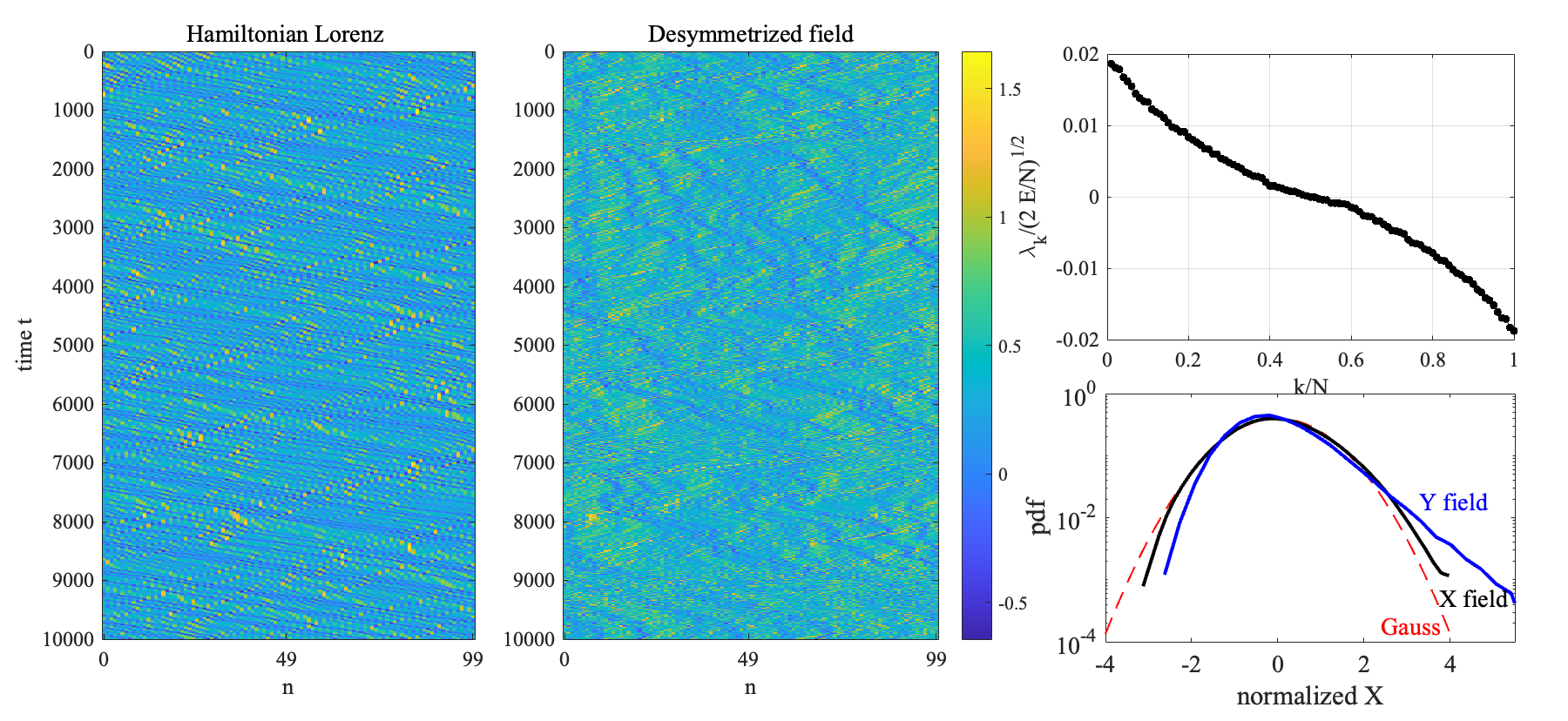}
    \caption{(Left panel) Space-time field $\{X_n(t)\}_{n=0,\ldots,N-1}$ of the Hamiltonian Lorenz-like model in Eq.~\eqref{eqn:eqnmot2}, with~$f(X)=1/(1+X^2)$, $N=100$ and Hamiltonian value~$E=25$ and Casimir invariants $C_0\approx 24.8$, $C_1\approx 22.7$; (Middle panel) desymmetrized field and (Right panel) Lyapunov spectrum and statistics of the thermalized states of the (black)~$\mathbf{X}$ and (blue)~$\mathbf{Y}$ fields. Their field variables are related by the transformation $Y_n=\phi(X_n)=X_n+X_n^3/3$.} 
    \label{Fig2}
\end{figure}

Next, we compare the dynamics of the Hamiltonian Lorenz and the original inviscid Lorenz-96 models with $N=100$. We define an initial random field~$\{\tilde{X}_n\}$ and normalize its amplitudes to have energy~$E=25$, that is~$\tilde{X}_n\rightarrow \tilde{X}_n\sqrt{2 E}/\sqrt{\sum_n \tilde{X}_n^2}$. For the specific case at hand, the associated Casimir invariants are $C_0\approx 24.8$ and $C_1\approx 22.7$. The left panel of Fig.~\ref{Fig2} depicts the corresponding space-time dynamics of the cubic Hamiltonian Lorenz-like model. This exhibits nonlinear advection masking the chaotic nature of the flow. This chaotic behavior is revealed in the desymmetrized field shown in the center panel, where the translation symmetry has been reduced using the first-Fourier slice approach~\citep{chaosbook,FedeleJFM2015}~(see also remark below). The associated Lyapunov spectrum is displayed in the right panel of the same figure. As expected, the spectrum is antisymmetric and indicates chaotic dynamics. Additionally, the Lyapunov spectrum, $\Lambda(E,C_0,C_1,N)$, is a function of the Hamiltonian, the two Casimir invariants and the value of $N$.  

\begin{figure}
    \centering
    \includegraphics[width=0.8\textwidth]{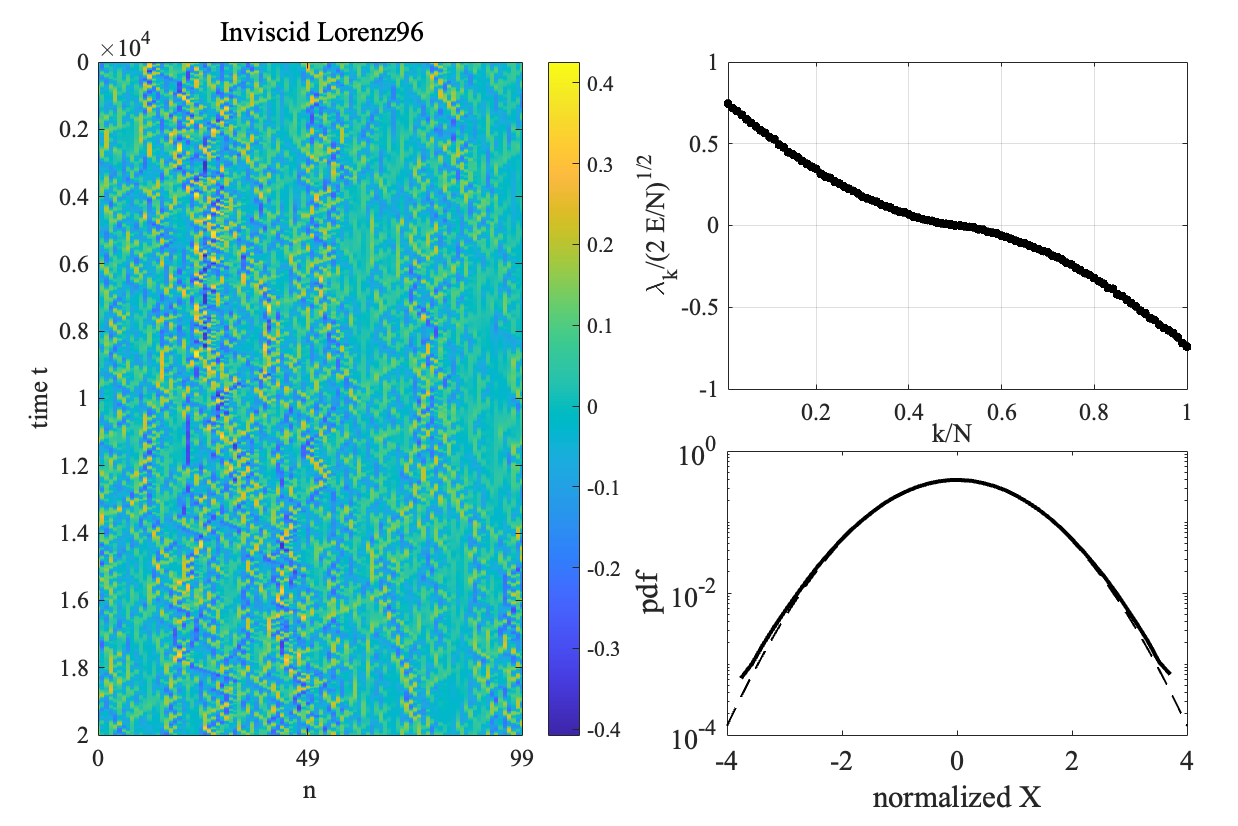}
    \caption{(Left panel) Space-time field of the inviscid Lorenz-96 model in Eq.~\eqref{eqn:eqnmot2}, $N=100$ and Hamiltonian value~$E=25$; (Right panel) Lyapunov spectrum and statistics of the thermalized state.} 
    \label{Fig3}
\end{figure}

For comparison, the space-time dynamics of the original inviscid Lorenz-96 model and associated Lyapunov spectrum are depicted in Fig.~\ref{Fig3}. The dynamics exhibit chaotic patterns, with no evidence of advection. Interestingly, the displayed Lyapunov spectrum is approximately anti-symmetric, but paired exponents differ at the second numerical digit. Since the model is homogeneous of second degree, the Lyapunov spectrum depends solely on the value $E$ of the Hamiltonian~$H$ and $N$. This implies the scaling~$\Lambda(E,N)=\sqrt{\frac{2 E}{N}}\Lambda(1,N)$, where $\Lambda(1,N)$ is the spectrum of the unit energy shell. 

\begin{remark}[Symmetry reduction]

The Lorenz models exhibit translation symmetry since their properties remain unchanged 
under a cyclic discrete shift in space. Reducing or quotienting out this symmetry allows for the study of the essential chaotic dynamics of the model, especially for high dimensional systems with ~$N\gg 1$~(see, e.g., \cite{Fedele_etalJFM2016}). For large~$N$, symmetry reduction helps removing the nonlinear advection observed in physical space~\cite{Fedele_etalJFM2016}~(see Fig.~\ref{Fig2}). To achieve this, we obtain a desymmetrized field~$X_n^{\rm des}$ by means of the first Fourier slice approach~\cite{chaosbook,Fedele_etalJFM2016}.

Given the discrete field ${X_n}$, the discrete Fourier transform (DFT) is defined as 
%------------------------------
$$
    \hat{X}_j = \sum_{n=0}^{N-1} X_n \exp{(-i k_j n)}\,,
$$
%------------------------------
where $j = 0, 1, 2, \dots, N-1$, and $\hat{X}_j$ is the Fourier coefficient corresponding to the spatial frequency, or wavenumber~$k_j=2\pi j/N$. Since ${X_n}$ is a real-valued field, then $\hat{X}_{-j}=\overline{\hat{X}}_j$, and $\hat{X}_0/N=\sum_n X_n/N$ is the mean value of the field. If the spatial labels of ${X_n}$ shift by $m$ the corresponding Fourier coefficients acquire the phase shift of $k_j m$. That is,
%------------------------------
$$
    \hat{X}_j\rightarrow \hat{X}_j \exp{(i k_j m)}\,.
$$
%------------------------------
We can now define the desymmetrized discrete field $X_n^{\rm des}$ through its Fourier components. These are obtained from~$\hat{X}_j$ by subtracting the generally time-varying phase~$\phi_1(t)$ of the first Fourier mode~$\hat{X}_1(t)=|\hat{X}_1(t)|\exp{[i \phi_1(t)]}$. Note that the phase must be unwrapped along time~\cite{Fedele_etalJFM2016}. The desymmetrized Fourier components are then given by
%------------------------------
$$
    \hat{X}_j^{\rm des}(t)=\hat{X}_j(t) \exp{[-i k_j\phi_1(t)]}\,.
$$
%------------------------------
These spectral coefficients are invariant under translation. Indeed, if the spatial labels of ${X_n}$ are shifted by $m$, then $\hat{X}_j\rightarrow \hat{X}_j \exp{(i k_j m)}$, and the phase of the first mode changes to~$\phi_1 + m$. As a result, the Fourier coefficients of the desymmetrized field transform as
$$
    \hat{X}_j^{\rm des}\rightarrow \hat{X}_j \exp{(i k_j m)}\exp{[-i k_j(\phi_1+m)]} = \hat{X}_j^{\rm des}\,,
$$
which means they remain unchanged.  

The desymmetrized field~$X_n^{\rm des}$ no longer exhibits translation symmetry, and the intrinsic chaotic dynamics of the system can be more easily analyzed, especially for high dimensional systems~$N\gg 1$. 
\end{remark}

%--------------------------------------
%--------------------------------------
%--------------------------------------
%--------------------------------------

%--------------------------------------
%--------------------------------------
%--------------------------------------
%--------------------------------------

\subsubsection{Statistics}

Our numerical experiments also indicate that for large $N$, the Hamiltonian Lorenz-like model thermalizes towards states with non-Gaussian statistics. The lower-right panel of Fig.~\ref{Fig2} depicts the probability density functions~(pdf) of the $\mathbf{X}$ and $\mathbf{Y}$ fields. Their field variables relate by the cubic transformation~$Y_n=\phi(X_n)=X_n+X_n^3/3$. As a result, the $\mathbf{Y}$-field is phase-locked to the $\mathbf{X}$-field, but with enhanced peak magnitudes or extremes. Consequently, the $\mathbf{Y}$-field exhibits heavier pdf tail than the $\mathbf{X}$-field. In contrast, the inviscid Lorenz-96 model thermalizes towards states with nearly Gaussian statistics, as seen in the lower-right panel of Fig.~\ref{Fig3}.

\section{Conclusions and Outlook}
\label{sec:conclusion}

While the Lorenz models have contributed fundamentally to our understanding of  chaotic dynamics and provided testbeds for new methods in weather forecasting and climate research, their non-Hamiltonian nature leads to nonphysical dissipation when representing fundamental physical principles. This nonphysical dissipation adds to the physical dissipation (mechanical damping), misleading its effects. 
Staying true to the original work by Lorenz~\citep{Lorenz2005}, we have designed new inviscid models that are both chaotic and Hamiltonian in nature, which we refer to as Hamiltonian Lorenz-like models. To achieve this, we started with a Poisson fluid bracket acting on continuous fields, known to satisfy the Jacobi identity. In doing so, we captured the Hamiltonian nature of the ideal part of the fundamental equations in fluid mechanics~\cite{Morrison1998}. We then discretized the bracket in a way that preserves the Jacobi identity. A family of Hamiltonian Lorenz-like models followed from a quadratic Hamiltonian. 
We have also developed a symplectic integrator to properly integrate these equations over long periods without affecting their fundamental properties, thereby avoiding numerical dissipation of the symplectic structure.

The new family of Hamiltonian Lorenz-like models we propose for geophysical fluid dynamics research, as a promising alternative to the original Lorenz models~\citep{Lorenz2005}, is given by
%----------------------------------------------------------
$$
    \dot{X}_n=f(X_n) \sum_{k=1}^K\xi_k \left( X_{n+k}f(X_{n+k})- X_{n-k}f(X_{n-k})\right)\,,
$$
%----------------------------------------------------------
where $\xi_k$ are arbitrary parameters, and $f(X)$ is a generic function that can be ad hoc chosen to qualitatively represent weather phenomena based on experimental or numerical data, or physical insights. For $K>1$, they resemble the Lorenz-2005 models with $K$-range coupling~\citep{Lorenz2005}. In addition to the Hamiltonian~$H=\sum_n X_n^2/2$, the dynamics conserves the Casimir invariant~$C=\sum_n \phi(X_n)$, where $\phi'(X)=1/f(X)$. 

For $K=1$ and $\xi_1=1$, we obtained a family of Hamiltonian models similar to the Lorenz-96 with nearest neighbor coupling as
%----------------------------------------------------------
$$
    \dot{X}_n=f(X_n)\left( X_{n+1}f(X_{n+1})- X_{n-1}f(X_{n-1})\right)\,.
$$
%----------------------------------------------------------
As an illustration, we selected one model from this family with $f(X)=1/(1+\alpha X + \beta X^2)$, where $\alpha$ and $\beta$ are real parameters such that $\beta> \alpha^2/4$. The Casimir invariant in this case is~$C=\sum_n X_n + \alpha X_n^2/2 + \beta X_n^3/3$. Numerical and theoretical investigations of this selected Hamiltonian Lorenz-like model revealed that it provides a qualitative representation of non-Gaussian weather extremes and wave interactions, which are key factors in understanding many climate system phenomena.

Fields where we foresee an especially useful application of this new family of Hamiltonian Lorenz-like models include data assimilation methods, predictability research, and climate statistics, the latter gaining increasing importance due to climate change~\citep[e.g.][]{LucariniSarno2011}. Recent efforts in high-resolution simulations of weather and climate, such as the European Union Destination Earth project~\cite{destination_earth}, involve kilometer-scale data assimilation, where nonlinearities are becoming a key issue~\citep[e.g.][]{Nerger2022}. The Hamiltonian reformulation of the Lorenz models offers an improved testbed for nonlinear data assimilation and predictability research.

\section*{Acknowledgments}

The authors would like to thank Predrag Cvitanovi\'c and Phil Morrison for useful discussions.

\section*{Data Availability Statement}  

The data and codes that support the findings of this study are available from the corresponding author upon reasonable request.

\section*{Author Contribution Statement}

{\bf F. Fedele}: Conceptualization, Methodology, Software, Writing - Original Draft. {\bf C. Chandre}: Methodology, Software, Writing - Original Draft. {\bf M. Horvat}: Methodology, Software, Writing - Review \& Editing. {\bf N. \v{Z}agar}: Writing - Review \& Editing.

%-----------------------------------------
%\bibliographystyle{elsarticle-num}
%\bibliography{refs}

\end{document}